\documentclass{article}
\usepackage{arxiv}
\usepackage{dcolumn}
\usepackage{mathptmx} 
\usepackage{multirow}
\usepackage{} 

\usepackage[utf8]{inputenc} 
\usepackage[T1]{fontenc}    
\usepackage{hyperref}       
\usepackage{url}            
\usepackage{booktabs}       
\usepackage{amsfonts}       
\usepackage{nicefrac}       
\usepackage{microtype}      
\usepackage{lipsum}
\usepackage{amssymb,amsmath,epsfig}
\newcommand{\beq}{\begin{equation}}
\newcommand{\eeq}{\end{equation}}
\newcommand{\bea}{\begin{eqnarray}}
\newcommand{\eea}{\end{eqnarray}}

\begin{document}
\begin{flushright}
AIP conference proceedings of the \\14th Asia-Pacific Physics Conference
\end{flushright}

\title{CIRCULAR GEODESICS IN KERR-NEWMAN-KASUYA BLACK HOLE}

\author{{Saeed Ullah Khan\thanks{saeedkhan.u@gmail.com}} \, and \, {Jingli Ren\thanks{Corresponding author: renjl@zzu.edu.cn}}\vspace{0.2cm} \\\vspace{0.08cm}
School of Mathematics and Statistics, Zhengzhou University, Zhengzhou 450001, China.}
\date{}
\maketitle
\begin{abstract}
This article explores the characteristics of ergoregion, horizons and circular geodesics around a Kerr-Newman-Kasuya black hole. We investigate the effect of spin and dyonic charge parameters on ergoregion, event horizon and static limit surface of the said black hole. We observed that both electric, as well as magnetic charge parameters, results in decreasing the radii of event horizon and static limit, whereas increasing the area of ergoregion. The obtained results are compared with that acquired from Kerr and Schwarzschild black holes. Moreover, we figured out the photons orbit of circular null geodesics and studied the angular velocity of a particle within ergoregion.\\
\\
\end{abstract}

\tableofcontents
\section{Introduction}
\label{sec:1}
Black holes (BHs), are among the most engaging and even strangest objects of our universe, due to their extremely dense spacetime region they possess much stronger gravitation attractions. The key property of a BH is to pull its surrounding objects, termed is the accretion \cite{Cunha}. Therefore, they are responsible for various interesting astrophysical phenomena: including but not limited to the formation of accretion disc; jets and gamma-ray bursts \cite{Berti}. The ergoregion of a BH has an essential consequence in astrophysics, as Hawking radiation could be analyzed in this region \cite{Hawking1}. The region is also momentous due to the occurrence of the Penrose process \cite{Penrose2}. The techniques used for energy extraction are of ample interest. Besides, the impact occurs within the ergoregion of a BH plays an essential role to understand the central mechanism of these processes \cite{Frolov}. A detailed analysis of the orbital circular motion around a Kerr-Newman (KN) BH, can be found in \cite{Puglies}. Pugliese and Quevedo \cite{Puglies1} by considering Kerr spacetime on the equatorial plane, investigated test particle motion. Khan et al. \cite{Saeed} studied the braneworld Kerr BH for the possible impact of the tidal charge on ergoregion and particle dynamics.

Since in the last few decades, particle dynamics around BHs have been an engaging problem, as they are of great interest and could convey eminent information and reveal the prolific structure of background geometry. Among various kinds of geodesics, the circular one is much engaging. They are very useful for understanding and explaining the quasinormal modes (QNMs) of a BH \cite{Nollert}. Chandrasekhar \cite{Chandrasekhar} was among the pioneers of geodesics investigators for Schwarzschild, Reissner-Nordström (RN) and Kerr BHs. Soon after his work, the study of geodesics caught a considerable interest of many researchers. Leiva et al. \cite{Leiva} by considering the rainbow gravity studied the geodesic of Schwarzschild BH. Cardoso et al. \cite{Cardoso} by exploring both charged and uncharged particles near a Kerr-Newman-Taub-NUT BH, explored all possible situation to the existence of circular orbits.
\section{Equatorial circular geodesics}
\label{sec:2}
This section is devoted to the exploring of particle dynamics near a Kerr-Newman-Kasuya (KNK) BH. For obtaining geodesics on the equatorial plan ($\theta=\pi/2$), we utilize the neutral particle motion. The corresponding space-time geometry to the exact solution of a rotating KNK BH, in Boyer-Lindquist coordinates, can be expressed as \cite{Kasuya} 
\begin{eqnarray}\label{K1}
ds^2& = &g_{tt}dt^2+2g_{t\phi}dt d\phi+g_{rr}dr^2+g_{\theta \theta}d\theta^2+g_{\phi \phi}d\phi^2,
\end{eqnarray}
with
\begin{eqnarray}\nonumber
&&g_{tt}=-\left(1-\frac{2Mr-e^2-Q^2}{\Sigma} \right), \quad g_{t\phi}= -a{\sin^{2}\theta}\left(\frac{2Mr-e^2-Q^2}{\Sigma}\right), \\\nonumber && 
g_{\phi \phi}=\left(\frac{(a^2+r^2)^2-\Delta a^2\sin^{2}\theta}{\Sigma}\right)\sin^{2}\theta, \quad g_{rr}=\frac{\Sigma}{\Delta}, \quad g_{\theta\theta}=\Sigma,
\end{eqnarray}
and $\Delta= r^2-2Mr +a^2 +e^2+Q^2, \,\, \Sigma=r^2+a^2 \cos^{2}\theta \,$. In which $M$, $e$ and $Q$, respectively represent mass, electric and magnetic charge, whereas $a$ denotes rotational parameter of the BH. The metric \eqref{K1}, reduces to the KN BH if $Q=0$; to the Kerr BH if $e=Q=0$; to the RN BH if $a=Q=0$; and to the Schwarzschild BH by letting $a=e=Q=0$.
\subsection{Ergoregion, horizons and the ergosurfaces}
\label{subsec:21}
Similar to the Kerr BH, the metric \eqref{K1} exactly have two horizons, namely the event and Cauchy horizon. In BH physics, the static limit surface is of considerable importance, as it changes the nature of a particle geodesics. We can easily obtain both of the horizons and ergosurfaces by solving $\Delta=0$ and $g_{tt}=0$, respectively. As a result
\begin{figure*} \vspace{0.0cm}
\includegraphics[width=0.95\textwidth]{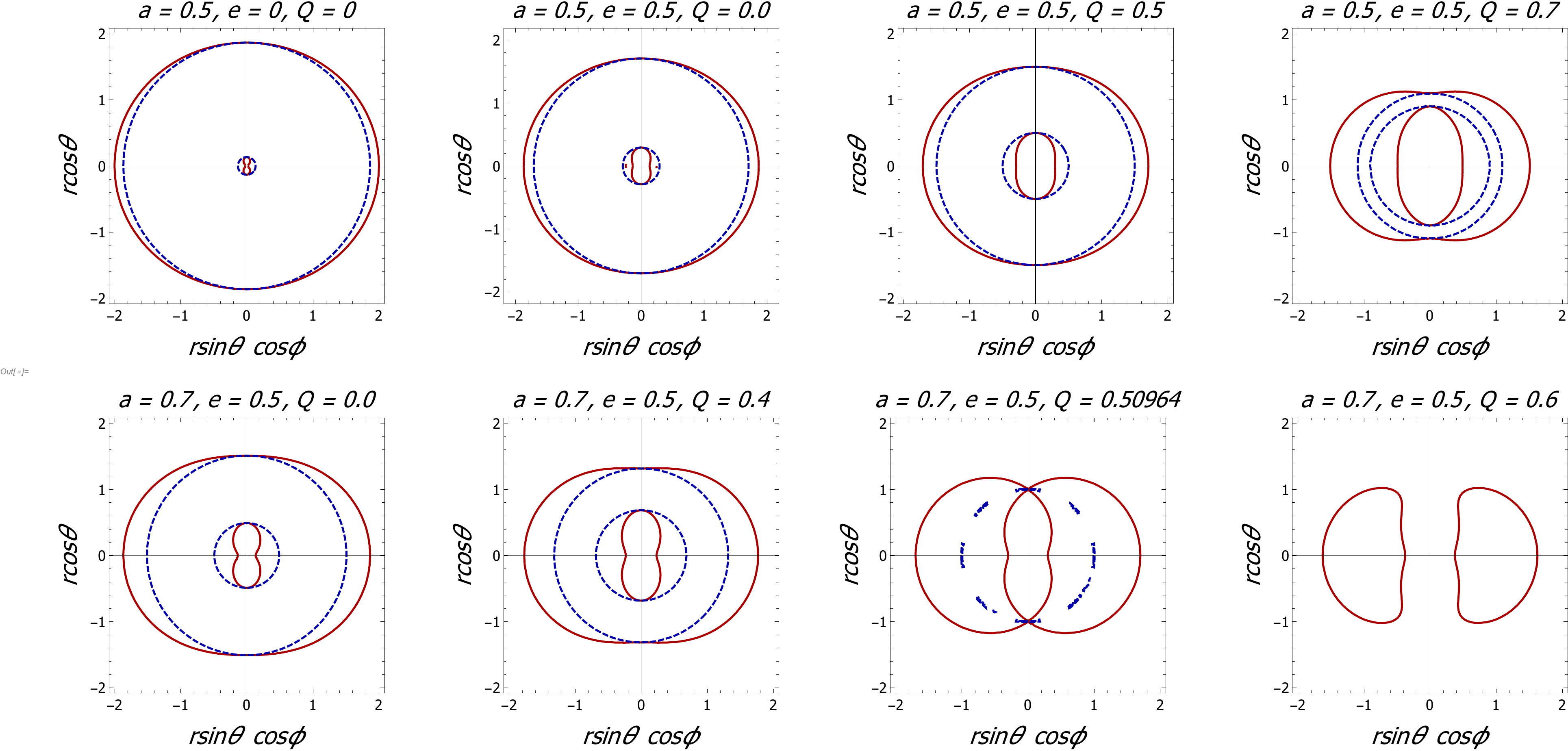}
\caption{Graphical description of ergoregion, event horizon (outer blue dotted curves) and static limit (outer red curves) in the xz-plane, at various points of the rotational and charge parameters $a$, $e$ and $Q$.}\label{ergo1}
\end{figure*}
\begin{eqnarray}\label{eh}
r_\pm &=&M \pm [ M^2-(a^2+e^2+Q^2)]^{1/2},
\\\label{sls}
r_{es\pm}&=&M \pm [ M^2-(a^2 \cos^2 \theta+e^2+Q^2)]^{1/2},
\end{eqnarray}
Equations \eqref{eh} and \eqref{sls}, should satisfy the constraints of $r_{es-} \le r_{-} \le r_{+} \le r_{es+}$. In rotating BHs, the region $\delta \equiv (r_{+}, r_{es+})$ is known as ergoregion. On considering BH spacetime, the $r$-coordinate is time-like in $r \in (r_{-}, r_{+})$, while space-like in  $r \in (0, r_{-}) \cup  r  > r_{+} $. This simply implies that the surfaces of constant $r$, i.e., $\delta_r$, are space-like for $\Delta < 0$, time-like for $\Delta > 0$ and light-like for $\Delta=0$ \cite{Penrose2}. In extreme KNK BH, i.e., for limiting value of the rotation parameter $a=\pm \sqrt {M^2-e^2-Q^2}$, horizons of the KNK BH coincides and for $cos^2\theta=1$, $r_{\pm}=r_{es\pm}$.
Figure \ref{ergo1}, reveals that both dyonic charge and spin of the BH has a considerable effect on the structure and size of ergoregion, horizons and static limit. Our investigation shows that increase in $a$, $e$ and $Q$ diminishing the radii of event horizon and static limit, whereas increases the area of ergoregion. One can also figure out the special cases with coinciding horizons and no horizons, as shown in  second row third and fourth columns, respectively. Moreover, we observed smaller radii of the event horizon and the static limit for KNK BH, but a greater area of the ergoregion in comparison with that of RN, Kerr and KN spacetime.
\subsection{Circular null geodesics}
\label{subsec:22}
By making use of the Lagrangian equation, the equation governing geodesics could be obtained as
\begin{equation}\label{Lagrangian}
\mathcal{L} = \frac{1}{2}g_{\mu\nu}\dot{x}^{\mu}\dot{x}^{\nu},
\end{equation}
here $\dot{x}^{\mu}={d x^{\mu}}/{d \tau}$ denotes the four-velocity with dot means $\partial / \partial \tau$ ($\tau$ is the proper time). 
Henceforth, by following \cite{Chandrasekhar}, the Lagrangian of KNK spacetime can be defined as
\begin{eqnarray}\label{eq2}
2\mathcal{L}&=&g_{tt}{\dot{t}^2}+2g_{t\phi} \dot{t}\dot{\phi}+ g_{\phi \phi} \dot{\phi^2} +g_{rr}{\dot{r}^2}.
\end{eqnarray}
Equation \eqref{eq2}, depicts that both $t$ and $\phi$ are cyclic coordinates with respect to the constants of motions, i.e., the angular momentum $L$ and total energy $E$, remains conserved along geodesics. Consequently, the corresponding generalized momenta take the form
\begin{eqnarray}\label{angmoment}
p_{t}&=&g_{tt}\dot{t}+ g_{t\phi}\dot{\phi}=-E,\\\label{energy}
p_{\phi}&=&g_{t\phi}\dot{t}+g_{\phi\phi}\dot{\phi}=L,\\
p_{r}&=&g_{rr}\dot{r}.\nonumber
\end{eqnarray}
In the case of KNK BH, the Hamiltonian will take the form
\begin{eqnarray}\label{hamiltonian}
2H=-\left(g_{tt}\dot{t}+g_{t\phi}\dot{\phi}\right)\dot{t}+\left(g_{t\phi}\dot{t}+g_{\phi \phi}\dot{\phi} \right)\dot{\phi} + g_{rr}\dot{r}^{2}
=-E\dot{t}+L\dot{\phi}+\frac{r^{2}}{\Delta}\dot{r}^{2}=\eta = constant,
\end{eqnarray}
different values of $\eta$ specify various types of geodesics, i.e., $\eta =-1,0,1 $ specify time-like, light-like and space-like geodesics, respectively. By making use of Eqs. \eqref{angmoment} and \eqref{energy}, one could obtain the radial equation of motion
\bea \label{radial}
\dot{r}^{2}=E^{2}+\frac{(a^{2}E^{2}-L^{2})}{r^{2}} +\frac{(aE-L)^{2}(2Mr-e^2-Q^2)}{r^4} + \frac{\Delta}{r^{2}}\eta .
\eea
The above result in Eq. \eqref{radial}, is of great interest, as it can be used to discuss the various associated features of particle dynamics. In case of null geodesics ($\eta=0$), the radial Eq. \eqref{radial}, could be expressed as
\begin{eqnarray}\label{ng1}
\dot{r}^2= E^2 + \frac{(a^{2}E^{2}-L^{2})}{r^{2}}  +\frac{(aE-L)^{2}(2Mr-e^2-Q^2)}{r^4}.
\end{eqnarray}
Following \cite{Shahzadi}, the equation of circular photons orbit can be acquired as 
\begin{eqnarray}\label{ng7}
r_{c}^3-3M r_{c}^2 +2(e^2+Q^2)r_{c} \pm 2a\sqrt{Mr_{c}^3-(e^2+Q^2)r_c^2}=0.
\end{eqnarray}
\begin{table*}[tbp]
\begin{center}
    \caption{Photon orbits $r_p$ for co-rotating particle.}
    \label{Tab1}
    \begin{tabular}{cc c c c c}
\hline \hline \noalign{\smallskip\smallskip}
& $e$ & ${r_p}$  &  {a} &  ${r_p}$  &\\ \noalign{\smallskip}
\hline \noalign{\smallskip\smallskip}
\multicolumn{1}{  c  }{\multirow{4}{*}{$Q=0$} } &
\multicolumn{1}{  c  }{ $0.0$ } & 2.493 & 0.0 & 2.823  \\ 
\multicolumn{1}{  c  }{}                        &
\multicolumn{1}{  c  }{ $0.3$ } & 2.421 & 0.2 & 2.568   \\
\multicolumn{1}{  c  }{}                        &
\multicolumn{1}{  c  }{ $0.5$ } & 2.281 & 0.4 & 2.281    \\
\hline \noalign{\smallskip\smallskip}
\end{tabular}%
    \begin{tabular}{cc c c c}
\hline \hline \noalign{\smallskip\smallskip}
& $e$ & ${r_p}$  &  {a} &  ${r_p}$    \\ \noalign{\smallskip}
\hline \noalign{\smallskip\smallskip}
\multicolumn{1}{  c  }{\multirow{4}{*}{$Q=0.5$} } &
\multicolumn{1}{  c  }{ $0.0$ } & 2.281  & 0.0 & 2.618  \\ 
\multicolumn{1}{  c  }{}                        &
\multicolumn{1}{  c  }{ $0.3$ } & 2.194 & 0.2 & 2.341   \\
\multicolumn{1}{  c  }{}                        &
\multicolumn{1}{  c  }{ $0.5$ } & 2.015 & 0.4 & 2.015    \\
\hline \noalign{\smallskip\smallskip}
\end{tabular}
\end{center}
\begin{center}\vspace{.5cm}
    \caption{Photon orbits $r_p$ for counter-rotating particle.}
    \label{Tab2}
    \begin{tabular}{cc c c c c c}
\hline \hline \noalign{\smallskip\smallskip}
& $e$ & ${r_p}$  &  {a} &  ${r_p}$ &\\ \noalign{\smallskip}
\hline \noalign{\smallskip\smallskip}
\multicolumn{1}{  c  }{\multirow{4}{*}{$Q=0$} } & 
\multicolumn{1}{  c  }{ $0.0$ } & 3.432 & 0.0 & 2.823  \\ 
\multicolumn{1}{  c  }{}                        &
\multicolumn{1}{  c  }{ $0.3$ } & 3.376 & 0.2 & 3.056   \\
\multicolumn{1}{  c  }{}                        &
\multicolumn{1}{  c  }{ $0.5$ } & 3.272 & 0.4 & 3.272    \\
\hline \noalign{\smallskip\smallskip}
\end{tabular}%
    \begin{tabular}{cc c c c}
\hline \hline \noalign{\smallskip\smallskip}
& $e$ & ${r_p}$  &  {a} &  ${r_p}$    \\ \noalign{\smallskip}
\hline \noalign{\smallskip\smallskip}
\multicolumn{1}{  c  }{\multirow{4}{*}{$Q=0.5$} } &
\multicolumn{1}{  c  }{ $0.0$ } & 3.272  & 0.0 & 2.618  \\ 
\multicolumn{1}{  c  }{}                        &
\multicolumn{1}{  c  }{ $0.3$ } & 3.210  & 0.2 & 2.866   \\
\multicolumn{1}{  c  }{}                        &
\multicolumn{1}{  c  }{ $0.5$ } & 3.0932  & 0.4 & 3.093    \\
\hline \noalign{\smallskip\smallskip}
\end{tabular}
\end{center}
\end{table*}
To figure out the radius of photons orbit, we assume that $r_{c}=r_{p}$, be the real positive root of Eq. \eqref{ng7}, which is the nearest possible photons orbit. By putting $Q=0$, $e=Q=0$ and $a=e=Q=0$, one could recovered the photons orbit of KN, Kerr and Schwarzschild BH, respectively. The behaviour of circular photon orbits is described in Tables \ref{Tab1} and \ref{Tab2}, respectively for both co-rotating and counter-rotating motion of particles.
\section{Particle's angular velocity within ergoregion}
\label{sec:3}
The current section is devoted to the study of a particle's angular velocity ($\Omega = d\phi/dt$) and its limitations in the vicinity of ergoregion under the restriction of $ds^2 \geq 0$. As a result
\beq\label{Ang1}
g_{tt}dt^2+2g_{t\phi }dt d\phi+g_{\phi \phi}d\phi^2 \geq 0,
\eeq
and it should satisfy the conditions of $\Omega_+\leq\Omega\leq \Omega_-$ \cite{Lightman}, here
\beq\label{Ang2}
\Omega_{\pm}={ (-g_{t\phi}\pm\sqrt{g_{t\phi}^2-g_{tt}g_{\phi\phi}}})/{g_{\phi \phi}}.
\eeq
\begin{figure*}
    \begin{minipage}[b]{0.57\textwidth} \hspace{0.3cm}
        \includegraphics[width=.75\textwidth]{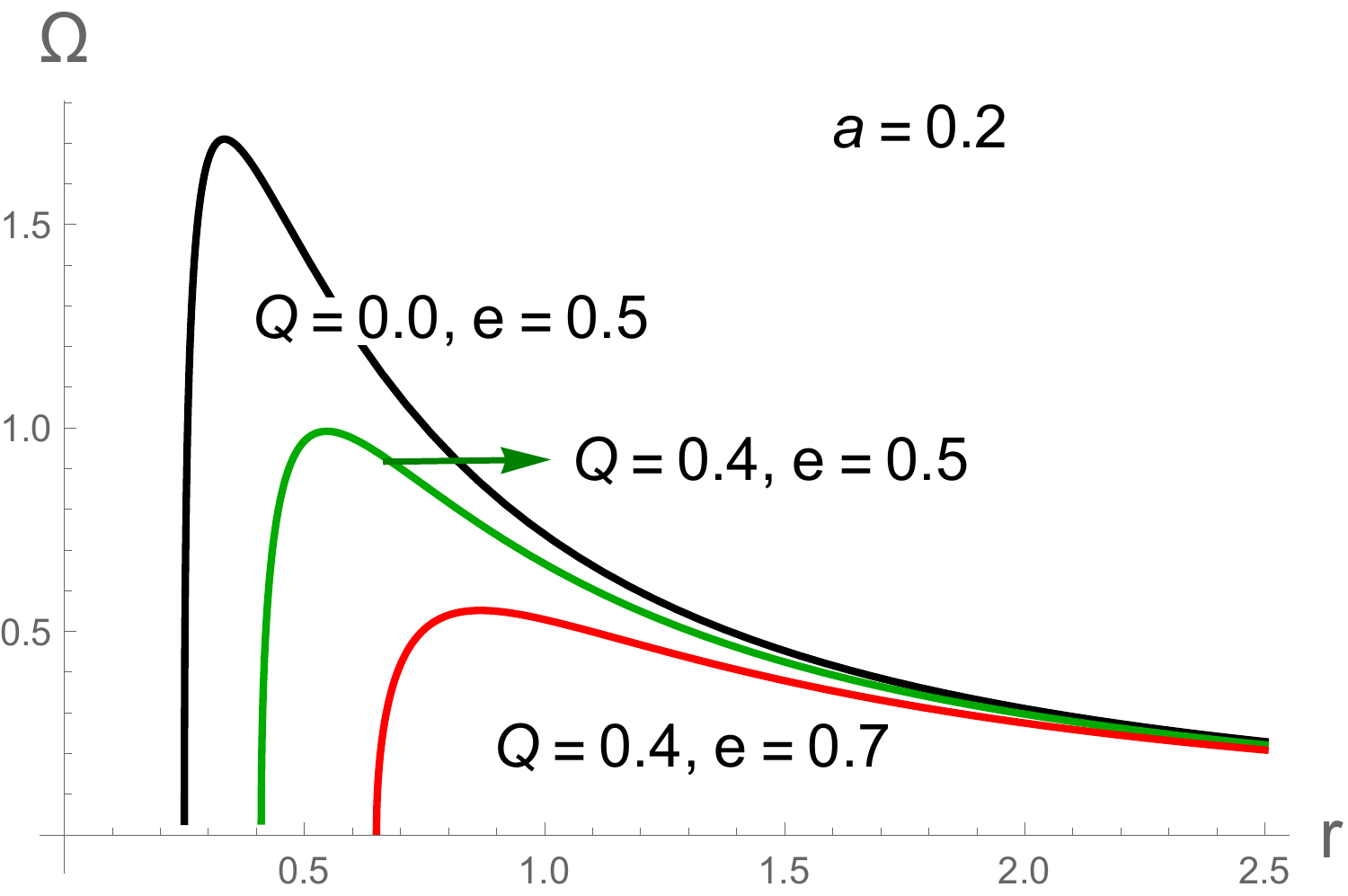}
    \end{minipage}\vspace{0.25cm}
        \begin{minipage}[b]{0.57\textwidth} \hspace{-0.6cm}
       \includegraphics[width=.75\textwidth]{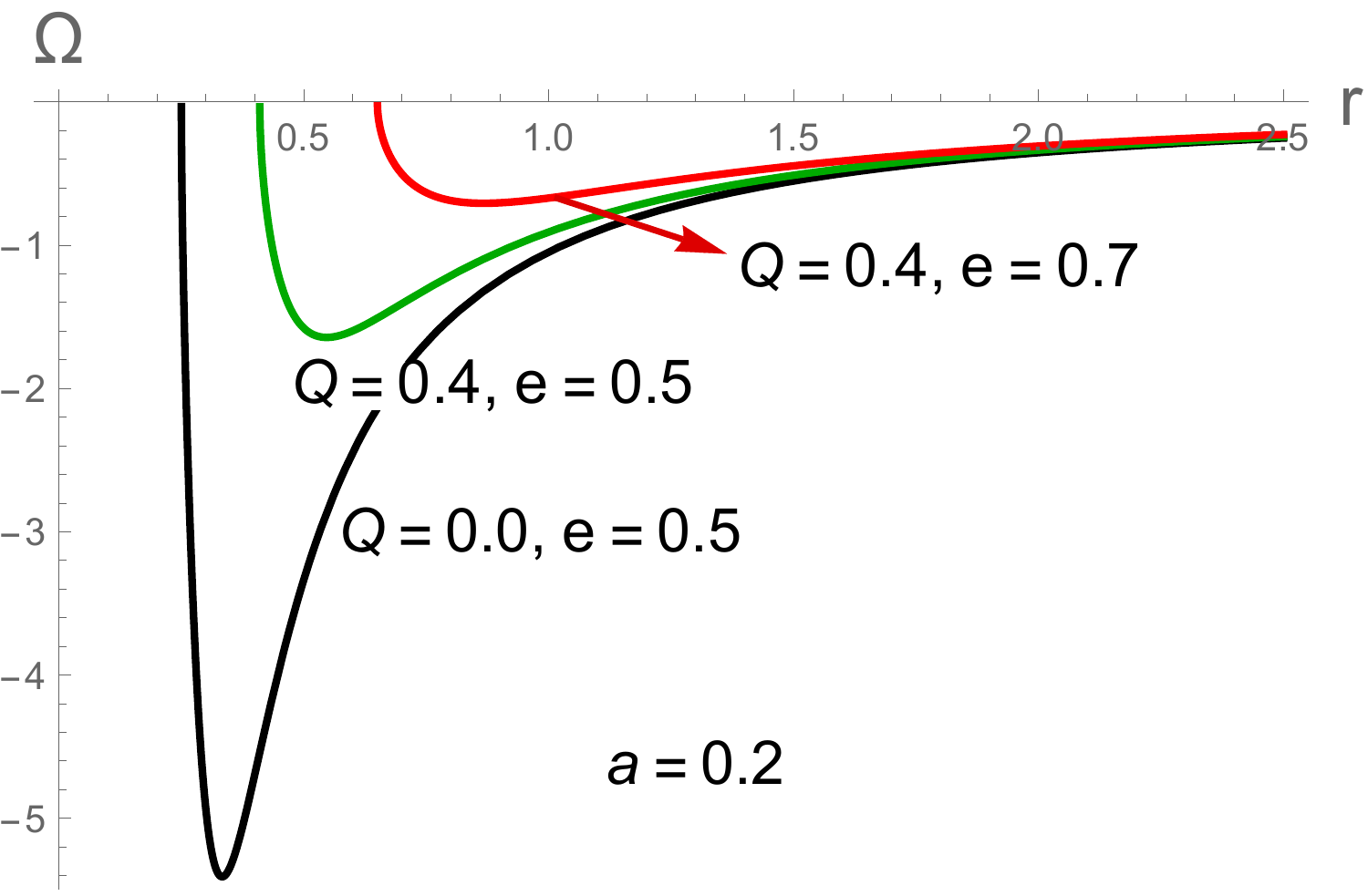}
    \end{minipage}
       \begin{minipage}[b]{0.57\textwidth}  \hspace{0.3cm}
        \includegraphics[width=.75\textwidth]{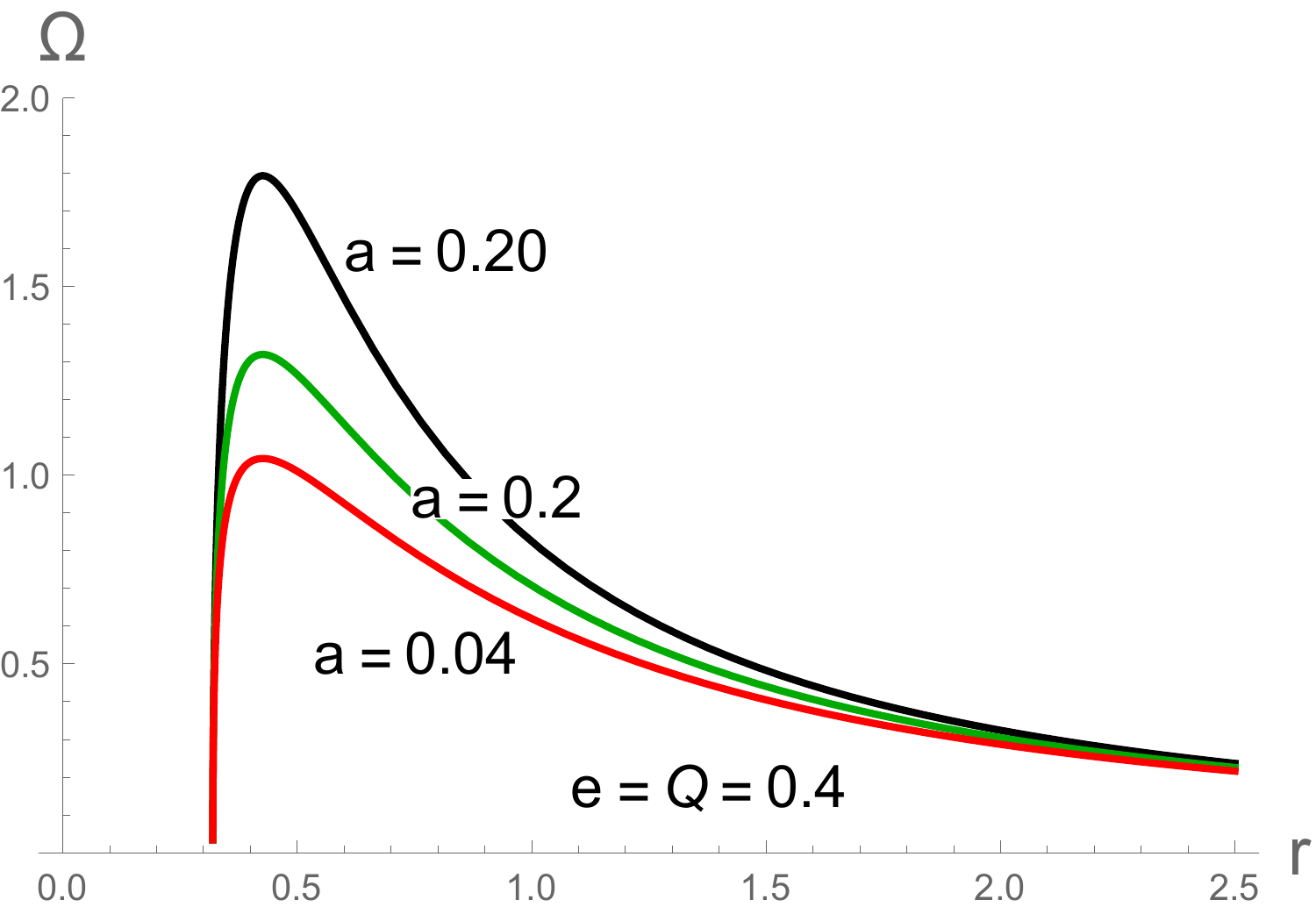}
    \end{minipage}
        \begin{minipage}[b]{0.57\textwidth} \hspace{-0.6cm}
       \includegraphics[width=.75\textwidth]{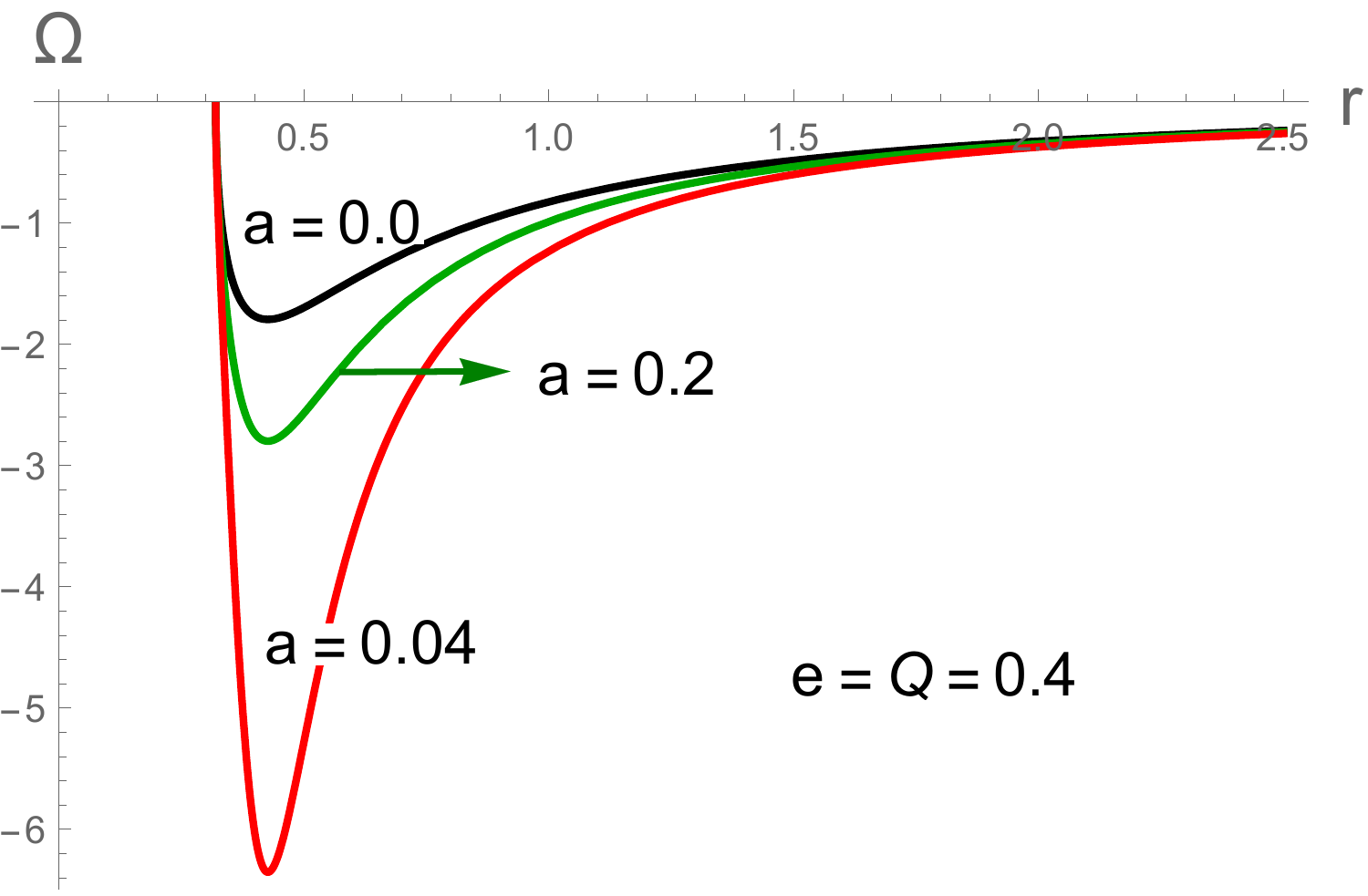}
    \end{minipage}
    \caption{Graphical interpretation of angular velocity $\Omega$, at different values of $a$, $e$ and $Q$ along with the radial distance $r$.}\label{Ang01}
\end{figure*}
While on ergosphere boundary, $g_{tt}$ vanished and $\Omega_+$ =0, whereas within ergosphere, $g_{tt}<0$ and $\Omega_{\pm}>0$, and each particle moves along BH rotation \cite{Chandrasekhar}. In the case of KNK spacetime Eq. \eqref{Ang2} simplifies to
\bea\label{Ang3}
&&\Omega_{\pm} = \omega \pm {\Sigma\sqrt{\Delta}}/{\sin{\theta}\left[(a^2+r^2)\Sigma+a^2 \sin^2{\theta}(2Mr-e^2-Q^2)\right]},
\eea
with
\begin{eqnarray}
\omega= {a\left(2Mr-e^2-Q^2\right)}/{\left[(a^2+r^2)\Sigma+a^2 \sin^2{\theta}\left(2Mr-e^2-Q^2\right)\right]}.
\end{eqnarray}
On approaching the event horizon, we get
\begin{equation}\label{Ang4}
\lim\limits_{r \to r_{+}} \Omega_{\pm} = 
\omega_{BH}={a(2Mr_{+}-e^2-Q^2)}/{\left[a^2 r_{+}^2 + r_{+}^4 + a^2 \left(2Mr_{+} -e^2-Q^2\right)\right]}.
\end{equation}
In the above equation, $\omega_{BH}$ is the angular velocity of BH rotation. Following \cite{Z.Stuchlik} the associated angular velocity of a particle at infinity turns out to be
\beq\label{Ang8}
\Omega={\pm\sqrt{Mr-e^2-Q^2}}/{(r^2\pm a\sqrt{Mr-e^2-Q^2})}.
\eeq
In Eq. \eqref{Ang8}, the upper and lower signs, respectively correspond to the co-rotating and counter-rotating orbits.
Figure \ref{Ang01}, describes the graphical interpretation of angular velocity $\Omega$, respectively at different values of dyonic charge and spin parameter $a$ of the BH. The graphical description reveals that angular velocity increases with the increase of dyonic charge, while decreases as BH rotation increases.
\section{Conclusion}
\label{sec:4}
In this manuscript, we carried out a detailed analysis of horizons structure and particles circular geodesic near a KNK BH. The obtained result ensure that both BH rotation and dyonic charge results in thicker the ergoregion, whereas decreasing the event horizon and static limit. As a result, the spacetime becomes denser. By making use of the radial equation of motion, we have derived the circular photons orbit. We observed that BH charge decreases the photons orbit of both direct and retrograde particles, while its spin diminishing the photons orbit of direct particles and contributes to the photons orbit of retrograde particles. Moreover, the angular velocity of a particle within ergoregion of a KNK spacetime has been explored. Our finding shows that angular velocity increases with dyonic charge, while decreases with BH rotation.

\end{document}